\documentclass[12pt,pre]{revtex4}

\usepackage{amssymb}
\usepackage{amsmath}

\newcommand{\pa}{\partial}

\newcommand{\myref}[1]{(\ref{#1})}

\newcommand{\de}{\delta}
\newcommand{\De}{\Delta}
\newcommand{\al}{\alpha}
\newcommand{\eps}{\varepsilon}
\newcommand{\la}{\lambda}
\newcommand{\La}{\Lambda}

\newcommand{\sig}{\sigma}

\renewcommand{\geq}{\geqslant}
\newcommand{\lan}{\langle}
\newcommand{\ran}{\rangle}

\newcommand{\demi}{\frac{1}{2}}

\newcommand{\sur}[2]{{\displaystyle\mathop{#1}_{#2}}}

\newcommand{\mcal}[1]{\mathcal{#1}}

\newlength{\somme}
\newlength{\sommep}
\newcommand{\intvide}{\rule[-\sommep]{0cm}{\somme}}

\newlength{\sommebis}
\newlength{\sommepbis}
\newcommand{\intvidepetit}{\rule[-\sommepbis]{0cm}{\sommebis}}

\usepackage{bm}
\usepackage{color}
\usepackage{graphics}

\newcommand{\mC}{\mcal{C}}

\begin{document}
\title{Injected Power Fluctuations in 1D dissipative systems : role of
  ballistic transport}
\author{Jean Farago}
\email[e-mail:]{farago@ics.u-strasbg.fr}
\affiliation{Institut Charles Sadron CNRS-UPR 22, 6 rue Boussingault BP 40016 F-67083 Strasbourg Cedex, France.}
\author{Estelle Pitard}
\email[e-mail:]{Estelle.PITARD@LCVN.univ-montp2.fr}
\affiliation{Laboratoire des Verres (CNRS-UMR 5587), CC69, Universit\'e Montpellier 2, 34095 Montpellier Cedex 5, France.}
\date{\today}
\begin{abstract}This paper is a generalization of the models considered in
  [\textit{J. Stat. Phys.}
  {\bf 128},1365 (2007)]. Using an analogy with free fermions, we compute exactly the large
  deviation function (ldf) of the energy injected up to time $t$
 in a one-dimensional
  dissipative system of classical spins, where a drift is
  allowed. The dynamics are $T=0$ \textit{asymmetric} Glauber
  dynamics driven out of rest by an injection mechanism, namely a
  Poissonian flipping of one  spin. The drift induces anisotropy in the system, making the model more comparable to
  experimental systems with dissipative structures. We discuss the physical content of
  the results, specifically the influence of the rate of the Poisson injection 
  process and the magnitude of the drift on the properties of the ldf. We also compare the results of this spin model to
  simple  phenomenological
  models of energy injection (Poisson or Bernoulli processes of domain wall injection). We show that many qualitative results
  of the spin model can be understood within this simplified framework.
\end{abstract}

\

\pacs{
02.50.-r Probability theory, stochastic processes, and
  statistics
05.40.-a Fluctuation phenomena, random processes, noise, and
  Brownian motion
05.50.+q Lattice theory and statistics (Ising, Potts, etc.)
}

\maketitle
\section{Introduction}

Dissipative systems are generically  systems for which a few relevant
degrees of freedom can be singled out and obey closed dynamical
equations: typically  a fluid, where the velocity field obeys the
Navier-Stokes equation, belongs to this category. Another well-known
example is given by  granular materials, where the identification of 
 relevant  variables (collisions) is even more evident. The lack of
completeness, caused by the selection of some degrees of freedom, gives however these systems a
nonconservative character, as  energy flows continuously from  relevant
degrees of freedom (kinetic energy) to  irrelevant ones (thermal
agitation). As a result, the dissipative systems are by nature very
different from the  systems usually suitable for the use of classical statistical
physics, where the conservation of energy is an unavoidable assumption.
 In particular, the whole set of tools devised by  statistical physics
can be of questionable use, even in  situations where  a
statistical approach seems natural: it is very tempting to interpret  turbulent
systems, or a vibrated granular matter, in terms of effective temperature,
correlations, Boltzmann factor, etc\ldots but the soundness of such an
approach is often questionable. 

Quite recently, the interest of physicists has been drawn to the
injection properties of dissipative systems for several
reasons. First, it was easily measurable experimentally, and the
measurements showed that, contrarily to what was usually expected, the injected
power fluctuates a lot, is not Gaussian, and does not obey the usual
simple scaling arguments \cite{fauveexp1,fauveexp2}. Moreover, the injection is by nature very
important in dissipative systems, since it is required to draw the
system out of rest; thus, it is natural to study specifically this
observable, which is at the same time  responsible for the existence of 
the stationary state, and is strongly affected by it \cite{aumaitrefauvemcnamarapoggi}. Finally, some theoretical works on the so-called
``Fluctuations Theorems'' had suggested a possible symmetry relation in the
distribution of the fluctuations of the injected power, a suggestion
vigourously debated since the works of \cite{aumaitrefauvemcnamarapoggi,evanscohenmorris,gallavotticohen,kurchan,kurchanrevue,cilibertolaroche}. 

In studying the fluctuations of global (macroscopic) variables of a
disordered (turbulent) dissipative system, one faces soon a crucial problem: contrary to
the statistical physics of conservative systems, no global theory is
at hand here to predict the level of fluctuations, the physical
meaning of their magnitude, the skewness of the distributions, etc\ldots
All these features are intimately connected to the statistical
stationary turbulent state, but in a way nowadays beyond our
knowledge. A way to make progress towards a better understanding of
these issues is to consider toy-models of dissipative systems where
some features of real systems are reproduced, and analyze the
structuration of the stationary states. If one can find for these
systems an intimate connection between their injection properties and
their dynamical features, such rationale could perhaps  be adapted  to
more realistic systems. Such a  procedure has been successfully applied in the
study of fluctuations of current for conservative systems \cite{bodineauderrida}.

In this paper, which follows  a former
one \cite{faragopitard}, we study a one-dimensional model of
dissipative system, which has the advantage to allow for an exact
description. This model consists of  a chain of spins subject to an {\it asymmetric} $T=0$ Glauber
dynamics, and is  driven out of rest
by a Poissonian flip of one spin (see next section for details): this is one of the rare examples where a nontrivial
stationary dissipative state can be entirely described. In fact, we generalized the symmetric model
studied in \cite{faragopitard} by allowing for an asymmetry in the diffusion dynamics. This system
is, in comparison with real dissipative systems, ridiculously
simple, but one can  hope that such examples would give ideas
to interpret real experiments or to explain measurements on other
variables, correlations, etc\ldots For that purpose, this paper
focuses much more on the physical content of the results than on the
computational details, that are postponed in the appendix.
More precisely, the observable we look at is the energy $\Pi$ provided to the system by the
injection mechanism between $t$ and $t+\tau$ in the permanent regime. For large
$\tau$, the probability distribution function (pdf) of $\Pi$ obeys the large deviation theorem and
the probability distribution function is  entirely governed by
the large deviation function $f$ (introduced below). The 
procedure of integrating the observable of interest over time has at least two advantages.
First, one can hope that this effective low-frequency filter fades
away ``irrelevant'' details of the dynamics and provides information on large-scale,
hopefully more universal phenomena at work; this statement has been proved
correct in some cases \cite{farago1,bodineauderrida}. Secondly, the experiments are
always constrained by a finite maximal frequency for the sampling of the time series: in practice 
the typical sampling time is much larger than 
the fastest relaxation times of the system
under consideration. As a result, the pdfs experimentally measured are
necessarily related to time integrated variables. The large deviation
function is a good representation  of these pdfs in the case where the sampling
frequency is small with respect to the dynamics of the bulk.

The paper is organized as follows: in the next section we define
precisely the
model; sections \ref{means} and \ref{s4} are devoted to the physical results given by our
computations. In section \ref{simplemodel}, we show that the main
characteristics of the large deviation function of the injected power
are  explained quite well using a simple phenomenological model, which
treats the correlations between the boundary and the bulk in an
effective way. The appendix (section \ref{calcul}) gives in detail all the steps of the
computation, based on a free-fermion approach of the intermediate
structure factor.

\section{The model}

We consider a 1D system of $N+3$ ($N\rightarrow\infty$) classical
spins on a line, labelled from $-1$ to $N+1$. The values of the extremal spins
$s_{-1}$ and $s_{N+1}$ are fixed (this choice makes the
description in terms of domain walls easier, as explained in the appendix). The zeroth
spin $s_0$ is the locus where  energy is injected into the system: the flipping
of $s_0$ is just a Poisson process with rate $\la$, independent of the
state of the other spins.  The spins of
the ``bulk'', from $s_1$ to $s_{N-1}$ are  updated
according to an \textit{asymmetric} $T=0$ Glauber dynamics. 

\begin{figure}[h]
   \centerline{\resizebox{9cm}{!}{\includegraphics{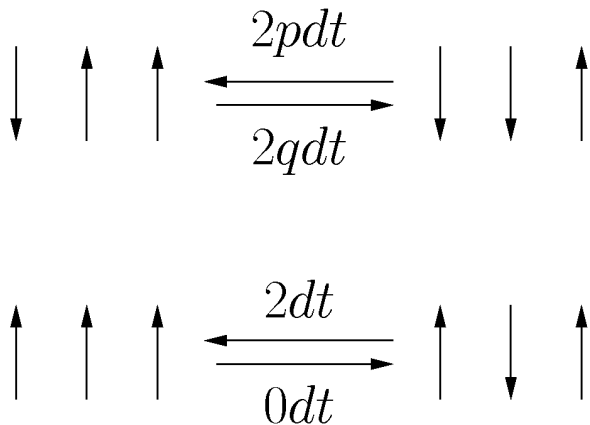}}}
\caption{The rates of the asymmetric Glauber dynamics}\label{xfig_flip}
 \end{figure}

The asymmetric $T=0$ Glauber dynamics is defined as follows : given
$0<p<1$, the probability for
a spin $s_j$ to flip between $t$ and $t+dt$ is 
\begin{align}
dt[1-s_j([1-p]s_{j-1}+ps_{j+1})],
\end{align}
 which is illustrated in figure \ref{xfig_flip}.
Note that if $p>q$ (resp. $<$), the
domain walls are locally drifted to the left (resp. right); for
$p=q=1/2$ we recover  the system studied in
\cite{faragopitard} (with the difference, that contrarily to \cite{faragopitard} the
system is not duplicated on each side of $s_0$;  this
simplification yields simpler calculations and a physics a
bit easier to analyze). Note that the case $p<1/2$, where the domain walls
easily invade  the system, is probably the most relevant one for a comparison with experimental devices of turbulent
convection.

These dynamics are
dissipative but a non trivial stationary state is nervertheless
reached thanks to the Poisson process on $s_0$ which injects
continuously energy into the system (the injected energy is positive on average;  however, negative energy
injections are also possible fluctuations due to the bulk dynamics).

\section{The mean injected power}\label{means}

\begin{figure}[h]
  \centerline{\resizebox{9cm}{!}{\includegraphics{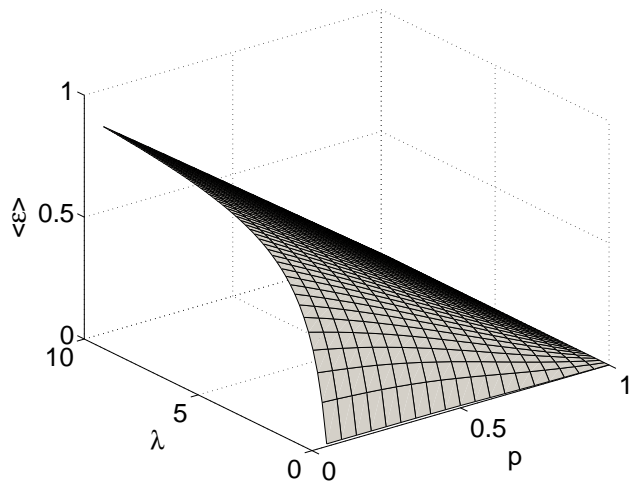}}}
\caption{Mean injected power as a function of $\lambda$ and $p$.}\label{m_i_p}
\end{figure}

The mean value of the injected power $\lan\eps\ran$ can easily be calculated. It is  given by
$\lan\eps\ran=\la[\text{Prob}(s_0=s_1)-\text{Prob}(s_0=-s_1)]=\la\lan
s_0 s_1\ran$. To compute $U_j=\lan s_0 s_j\ran$ (we are interested here in the special case  $j=1$), we
notice that the quantity $U_j$ obeys a closed equation in the permanent regime
(see \cite{farago1} for details):
\begin{align}
  -(\la+1)U_j+pU_{j+1}+qU_{j-1}&=0
\end{align}
(let us recall that $q=1-p$) with the boundary conditions $U_0=1$ and $U_\infty=0$. The
determination of $U_j$ is simple : the
polynomial $pX^2-X(\la+1)+q$  has a unique root $r$ less than
one, and therefore $U_j=r^j$. The mean injected power 
\begin{align}\label{Imoy}
\lan  \eps\ran&=\la\frac{\la+1-\sqrt{(\la+1)^2-4pq}}{2p}
\end{align}
is plotted in figure
\ref{m_i_p}. One can see that it
is an increasing function of  $\la$ and a decreasing function of
$p$. This last point can be  easily understood, as for higher $p$ 
the domain walls are more and more confined near the boundary, enhancing the
probability of negative energy injection. On the contrary, for low
values of $p$, the domain walls invade the system rather easily: for $p<1/2$, they are drifted away from the site of injection.
This
is at the origin of a large positive value for the average injection energy.

\section{The large deviations of the injected power}\label{s4}

The main result of our paper is the  computation of 
$f(\eps)$, the large
deviation function of the injected energy. This is  a central observable
associated with the long time (or low frequency) properties of the  fluctuations of the energy flux in stationary
systems. Let us call $\Pi$ the energy injected into the system between
$t=0$ and $t=\tau$; typically $\Pi$ scales like $\tau$ for large $\tau$. The ldf $f(\tau)$ is defined as 
\begin{align}
  f(\eps)=\lim_{\tau\rightarrow\infty}\tau^{-1}{\ln [ \text{Prob}(\Pi/\tau=\eps)]}
\end{align}
However it is simply defined, this quantity
is difficult to compute or analyze theoretically, as it involves the
 knowledge of the complete dynamics of the system, and measures the
temporal correlations which develop in a nontrivial way in the
nonequilibrium stationary state.

Usually, one computes first the ldf $g(\al)$ associated with the
generating function of $\Pi$:
\begin{align}
\lan e^{\al \Pi}\ran \sur{\simeq}{\tau\rightarrow\infty}e^{\tau g(\al)}
\end{align}

More precisely $g(\al)=\lim_{\tau\rightarrow\infty} \tau^{-1}\log\lan e^{\al \Pi}\ran$.
 Then,  $f(\eps)$ can be obtained  numerically
 solving the inverse Legendre transform
 \begin{align}
   f(\eps)&=\min_{\al}\left(g(\al)-\al \eps\right)
 \end{align}
 
The details of the computation of $g(\al)$ are postponed in the
Appendix. The formula for $g(\al)$ (equation
\myref{finalformulaforg}) is not easy to interpret physically. We are
thus in a  situation where the exact result does not
really highlight the underlying physics, and in particular does not
make  the long-time properties of the injection
process particularly transparent. In order to clarify this, we will
follow a very pragmatic way: first we will sketch the different ldfs corresponding to
different values of the relevant parameters $(\la,p)$ and raise some questions associated to
them. In the next section, we will see that some simple
phenomenological models account very well for the observed behaviours
(these models were neither discussed nor even evoked in \cite{faragopitard}).

\begin{figure}[h]
  \centerline{\resizebox{13cm}{!}{\includegraphics{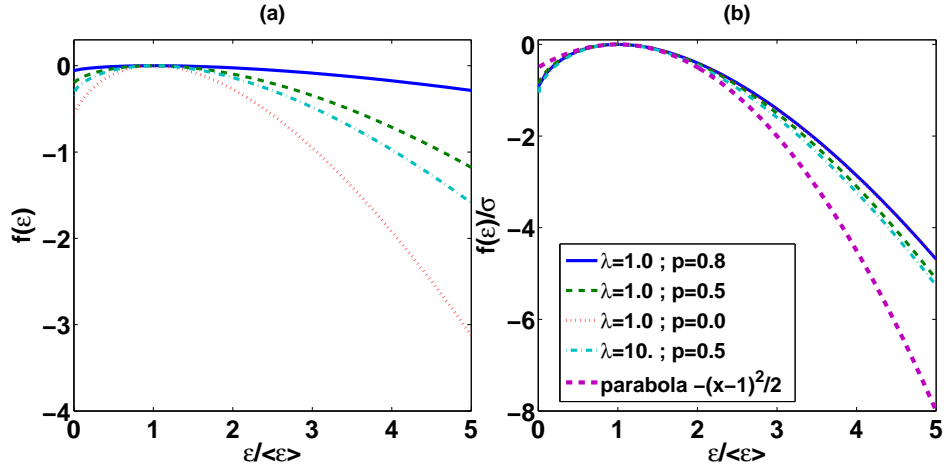}}}
\caption{(Color online) Large deviations functions $f(\eps)$ as a function of
  $\eps/\lan\eps\ran$ for various values of the parameters $p$ and
  $\la$. (a) curvature not rescaled, (b) curvature normalized}\label{ldf_vrac+rescaled}
\end{figure}

\begin{figure}[h]
  \centerline{\resizebox{13cm}{!}{\includegraphics{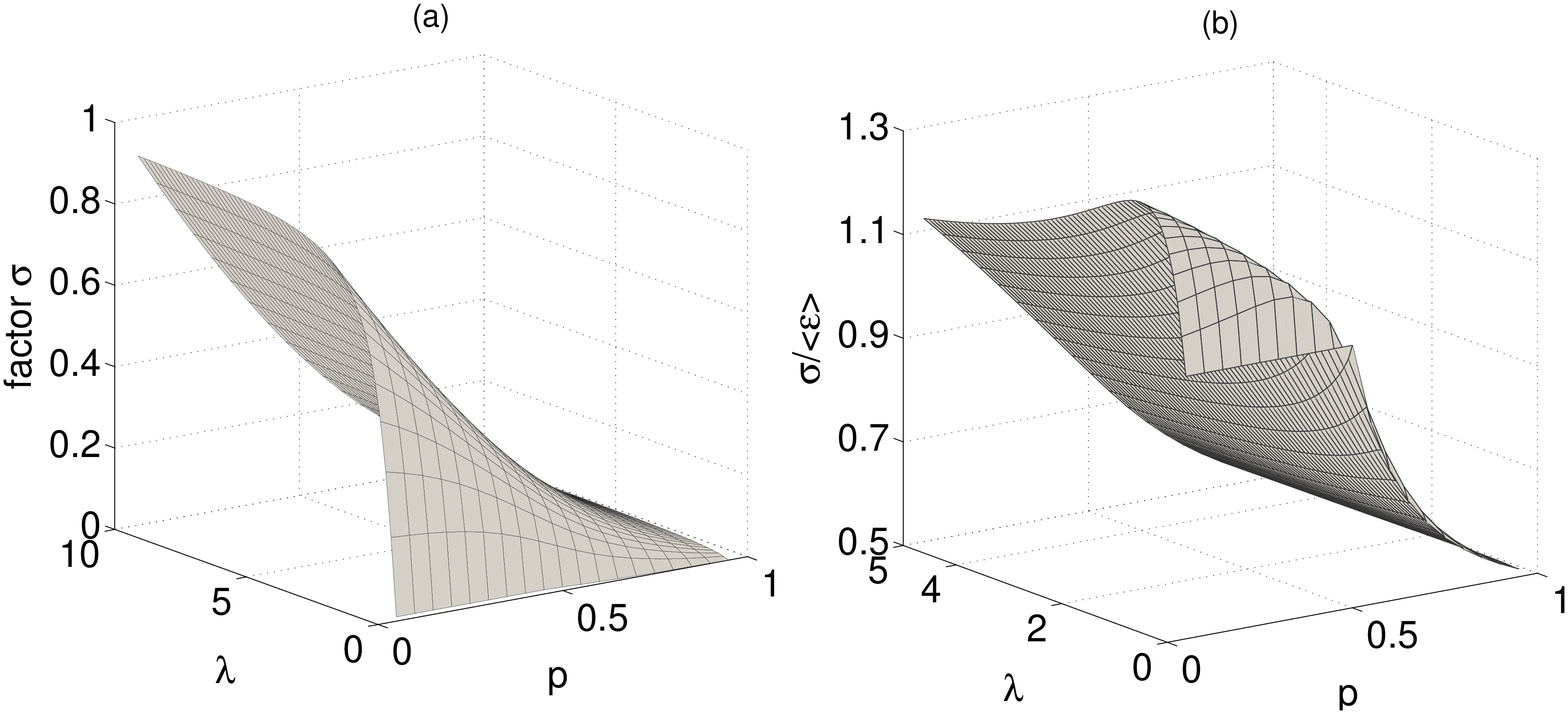}}}
\caption{(a) Factor $\sigma=-f''(\lan\eps\ran)\lan\eps\ran^2$ and (b)
  Fano factor $\sig/\lan\eps\ran$, as a
  function of $\la$ and $p$.}\label{sigma+Fano}
\end{figure}

In figure \ref{ldf_vrac+rescaled} (a), we show various functions $f(\eps)$ for different 
values of the parameters $\la$ and $p$, as a function of $\eps/\lan\eps\ran$. $f(\eps)$ is maximum 
for $\eps=\lan\eps\ran$, which is a generic property of ldfs.
Clearly, the curvature at the maximum is a major feature of these
curves, and is strongly dependent on the parameters $(\la,p)$. Writing $f(\eps)=-\demi
      [-f''(\lan\eps\ran)\lan\eps\ran^2](\eps/\lan\eps\ran-1)^2+o(\eps/\lan\eps\ran-1)^2$, we see that the relevant quantity 
      associated to the curvature, once $\eps$ has been rescaled by $\lan\eps\ran$,
is
$\sigma=-f''(\lan\eps\ran)\lan\eps\ran^2=g'(0)^2/g''(0)$. The
curves rescaled by the curvature $\sig$ are plotted in figure
\ref{ldf_vrac+rescaled} (b), where it is seen that the
curvature and the mean energy, though of  primordial importance,  are however not sufficient to
characterize fully the ldf: there is no clear collapse of the curves. The
dependence of $\sigma$ with respect to $p$ and $\la$ is plotted
in figure \ref{sigma+Fano} (a).
Its behaviour  is remarkably
similar to that of $\lan\eps\ran$ itself. 
Before explaining this point (see next section), we note that the mean value of the energy injected up to time $\tau$
is $\lan\Pi\ran=\tau\lan \eps\ran$; besides, the (squared) relative fluctuations
of this quantity is given for large $\tau$ by
$[\lan\Pi^2\ran-\lan\Pi\ran^2]/\lan\Pi\ran^2= 1/(\tau\sig)$. The
ratio of these two quantities
$[\lan\Pi^2\ran-\lan\Pi\ran^2]/\lan\Pi\ran=\sig/\lan\eps\ran$, 
called the Fano factor, is plotted in figure \ref{sigma+Fano} (b):
it
is comprised between 0.5 and 1.3 for all values of the parameters
$(\la,p)$, which shows that the correlation between $\sig$ and $\lan \eps\ran$, though clearly
demonstrated, is a bit loose. Thus, a sound question is
to ask why $\sig$ and $\lan \eps\ran$ are correlated, and what are the
factors which limit or modulate this correlation. These issues
will be discussed in the next section.

\begin{figure}[h]
   \centerline{\resizebox{16cm}{!}{\includegraphics{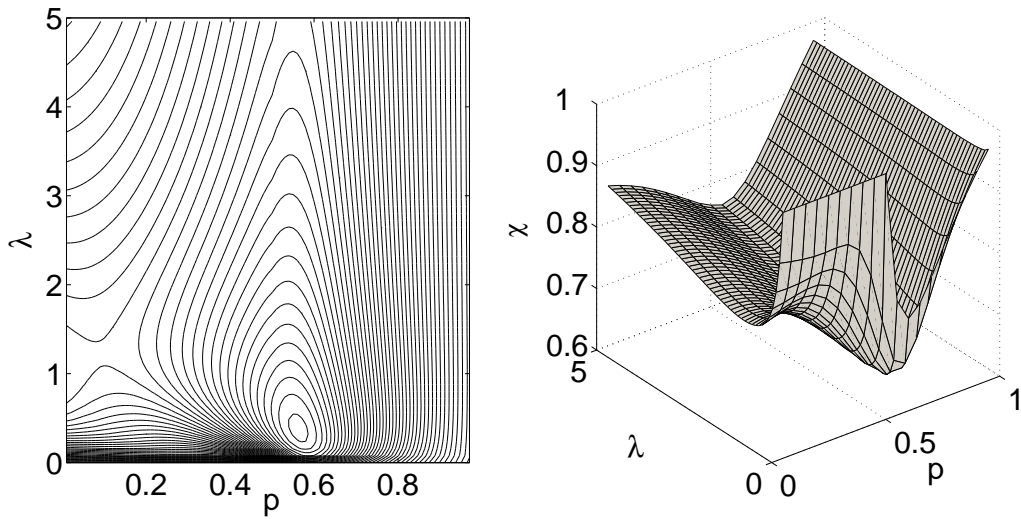}}}
\caption{Parameter $\chi$ as a function of $p$ and $\la$. The left
  graph is a contour plot of the surface.}\label{chi}
\end{figure}

Let us go back to the rescaled ldf in figure \ref{ldf_vrac+rescaled}
(b). One can notice that all curves display a noticeable counterclockwise
tilt with respect to the parabola. As for the Fano factor, this
tilt seems to be constant,  with some minor relative
differences. To quantify this tilt, one writes the Taylor
expansion of $f(\eps)/\sig$ up to the third order like:
\begin{align}
f(\eps)/\sig=-\demi(\eps/\lan\eps\ran-1)^2+\frac{\chi}{6}(\eps/\lan\eps\ran-1)^3  +o(\eps/\lan\eps\ran-1)^3
\end{align}
A simple calculation gives $\chi=g'''(0)g'(0)/g''(0)^2$. This
parameter quantifies  the tilt and can be a priori positive
or negative. The variation of $\chi$ with $(p,\la)$, plotted in
figure  \ref{chi},
shows a rather complicated dependence of $\chi$ with respect to the
parameters (in particular an absolute minimum for $p\sim 0.6$ and
$\la\gtrsim 0.5$), but with always $\chi\in[0.6,1]$. Both the global
trend and the finer details raise natural questions: why is the
tilt is always positive ? What  does it mean concerning the physics of the
system~? Why is the dependence on $p$ and
$\la$ so complicated~? The next section provides a phenomenological model which
give satisfactory answers to these issues .

\section{Discussion and comparison with a simplified model}\label{simplemodel}

In this section we  compare the  results obtained above to a very simple model, in order to see which 
global mechanisms are at work.

We consider an oversimplified version of our system, called in the following
``pure Poissonian model'' or PPM. In this model, the injection is
 a Poissonian emission (with rate $\rho$) of domain walls (d.w.) into the system and the energy is
incremented by one each time a d.w. is emitted. In this case, on has
$\lan\Pi\ran=\tau\rho$ and
$[\lan\Pi^2\ran-\lan\Pi\ran^2]/\lan\Pi\ran^2=1/(\tau\rho)$, that is
$\lan\eps\ran$ and $\sig$
are both equal to $\rho$. Thus, in our system, the global similarity between the
two quantities, i.e. $\sig/\lan\eps\ran\sim 1$ is not fortuitous, it
is in fact a signature of the approximate Poissonian
structure of the injection.

 Conversely, the violation of the relation
$\sig/\lan\eps\ran=1$, plotted in figure \ref{sigma+Fano} (b), is interesting, as it accounts directly for the
coupling of the Poissonian injection and the structuration of the
system near the boundary. 
In order to clarify this coupling, we can extend slightly the PPM to account
for the variability of the Fano factor, by considering that in an
``effective'' energy injection, not only one domain wall is concerned,
but in fact an average number of domain walls $n_\text{dw}$. For instance, for $p\simeq 1$,
where the domain walls are confined at the boundary,  the
only way for the system to absorb energy is the following rare
event: a domain wall is created between $s_0$ and $s_1$, it translates to the right
(limiting factor), and then comes back to annihilate with another
entering domain wall. This ``injection event'' is so rare, that two
such events are necessarily far apart from each other, and the statistics of these events is actually
Poissonian. In fact, one can see that the effective number of domain walls
associated to one event is two instead of one. Indeed, if one
generalizes the PPM to emit $n_\text{dw}$ domain walls per event, one
gets $\sig/\lan\eps\ran=1/n_\text{dw}$: figure \ref{sigma+Fano} (b)
gives $n_\text{dw}\simeq 2$ in the $p\simeq 1$ region as
expected. 

Another region is simple to analyze: for $\la\simeq 0,
p<0.5$ (the domain walls invade the bulk), the inner dynamics is so slow that the effective emission of
domain walls invading the bulk is Poissonian; virtually no
domain wall is reabsorbed by the boundary. One understands that this
scenario breaks down rather abruptly when the drift is directed towards
$s_0$, which explains the singular behaviour of the Fano factor at
$(\la,p)=(0,0.5)$.
To summarize, the fact that $\sig/\lan\eps\ran<1$ for the most
part of the parameter range illustrates the cooperative character of
the energy injection.

However, in the special case of  large $\la$, small $p$, one
also expects  a Poissonian behaviour, determined in this case by the natural
time of the bulk dynamics: for very quick flipping of the spin $s_0$,
and $p\gtrsim 0$, the emission of a domain wall into the system is only
limited by the move of the first domain wall  to the right.

The PPM, even modified by the parameter $n_\text{dw}$ is 
unable to account for the region where the Fano factor is larger than
one (namely this regime of large $\la$, small $p$): it is difficult to imagine an effective Poissonian emission of an
average number of domain walls less than one. Moreover, if one
also considers also the tilt parameter $\chi$, the disagreements are
stronger, for it can be easily shown that the PPM (with $n_\text{dw}$
allowed) yields $\chi=1$, irrespective of the value of $n_\text{dw}$.
We conclude that if the global trend of a positive tilt is again a
signature of the approximate Poissonian nature of the domain wall
injection, the model is a bit too rough to account for the observed
subtleties (except for regions where $\chi\simeq 1$, which correspond
to the cases commented  above).

\medskip

In order to get a finer description of the phenomenology, we can  add a new parameter in the PPM model.
 Instead of assuming a Poisson process for the emission of
 domain walls, we assume a Bernoulli process \cite{bernoulli}: the time span $[0,t]$
is divided into $t/\De t$ intervals of length $\De t$, during which
$n_\text{dw}$ domain walls can be emitted with a probability $\rho\De
t$. One thus takes into account a possible waiting time after an
emission event during which no other event is on average allowed.
For this model, one easily shows that
\begin{align}
  \lan\eps\ran&=\rho n_\text{dw}\\
\sig/\lan\eps\ran&=\frac{1}{n_\text{dw}(1-\rho\De t)}\\
\chi&=\frac{1-2\rho\De t}{1-\rho\De t}
\end{align}
We remark that now the Fano factor can reach values less than
one. This is the case for small values of $\la$ and $p<1/2$, where
$n_\text{dw}$ is certainly one: here the Poissonian character of the process
is imposed by $\la$, but there can be a waiting period after a flipping
of $\la$, due to the finite time required for the bulk dynamics to
remove the domain wall from its first position.

We remark also that the $\chi$ factor of the Bernoulli model is always less than one, exactly
like in the real system. It confirms also our previous interpretation for
the case $\la\simeq 0$ and $p<1/2$: a deep decrease of $\chi$ is
observed for increasing values of $\la$, which is associated in the
Bernoulli model with an increase of $\De t$. By the way, we can
extract from the preceding equations the effective
parameters  $n_\text{dw}$, $\De t$ and $\rho$, knowing $\lan\eps\ran$,  $\sig$ and $\chi$ from our numerical computation:
\begin{align}
   \De t&=\frac{1-\chi}{\sig}\\
n_\text{dw}&=(\sig/\lan\eps\ran)^{-1}(2-\chi)\\
\rho&=\frac{\sig}{2-\chi}
\end{align}
\begin{figure}[h]
   \centerline{\resizebox{8cm}{!}{\includegraphics{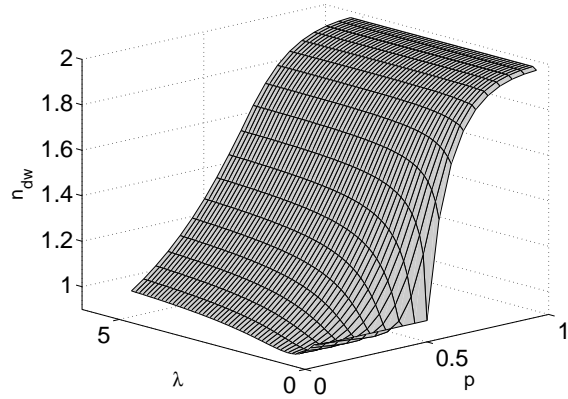}}}
\caption{Effective parameter $n_\text{dw}$ as a function of $p$ and $\la$.}\label{ndw}
\end{figure}
\begin{figure}[h]
   \centerline{\resizebox{8cm}{!}{\includegraphics{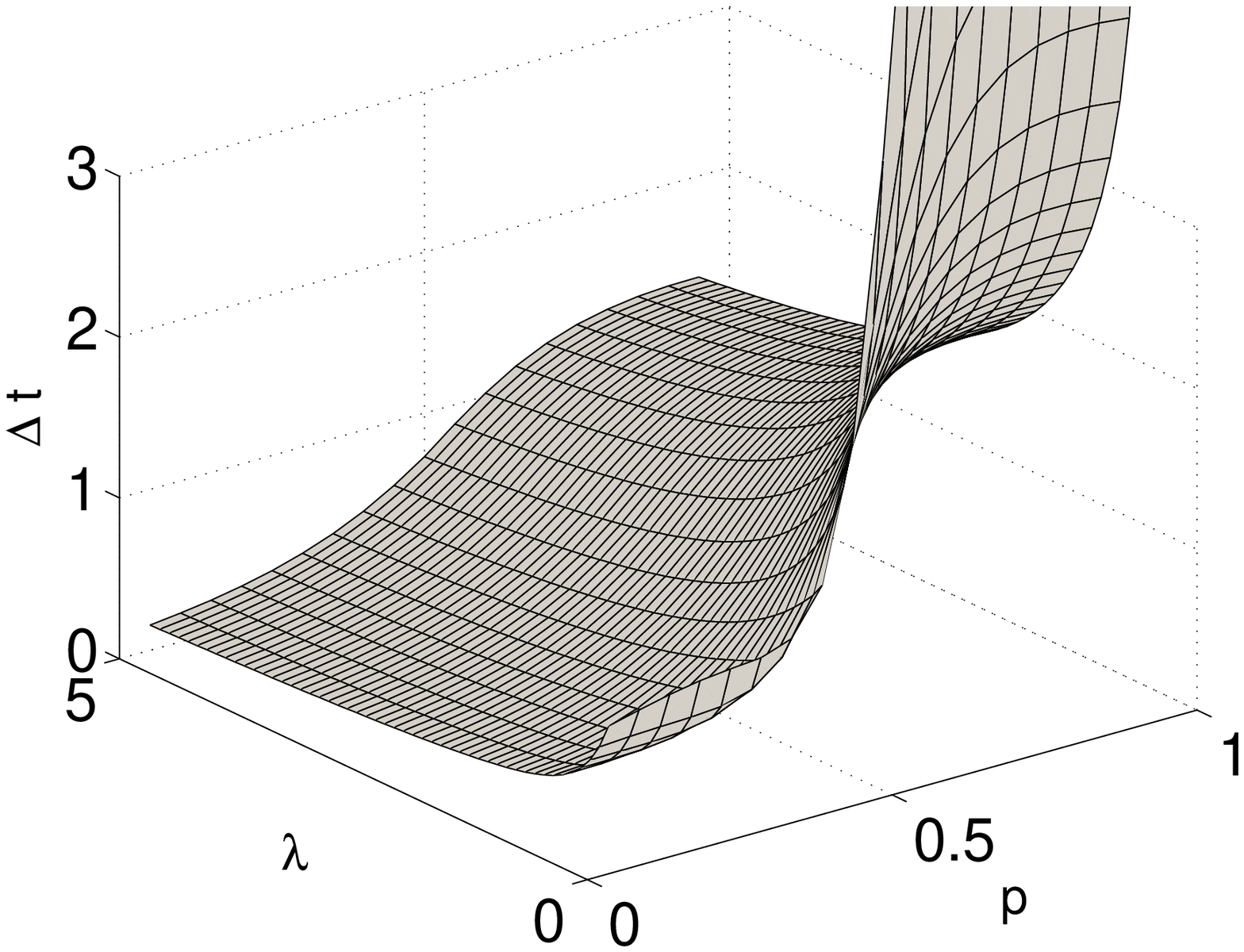}}}
\caption{Effective parameter $\De t$ as a function of $p$ and $\la$.}\label{deltat}
\end{figure}
\begin{figure}[h]
   \centerline{\resizebox{16cm}{!}{\includegraphics{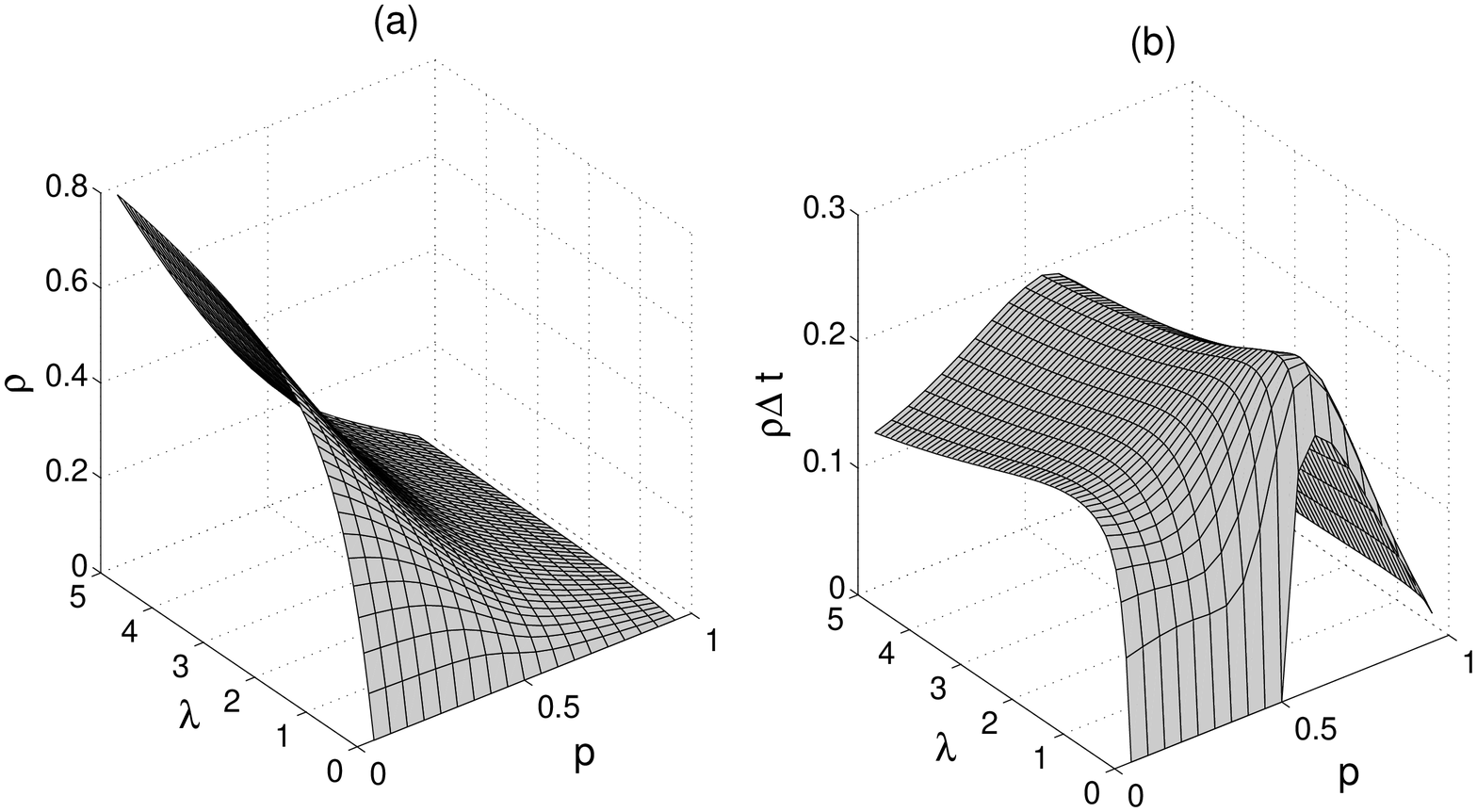}}}
\caption{Effective parameters (a) $\rho$ and (b) $\rho\De t$ as a function of $p$ and $\la$.}\label{rhoandrhodeltat}
\end{figure}
In figure  \ref{ndw},  \ref{deltat}, \ref{rhoandrhodeltat} (a) and
\ref{rhoandrhodeltat} (b), we see the values of
$n_\text{dw}$, $\Delta t$, $\rho$ and $\rho\Delta t$ respectively, extracted from the results for the spin model.
 The interpretation of figure \ref{ndw} is obvious: as
expected, the average number of domain walls $n_\text{dw}$
\textit{stays close to one} for $p\gtrsim 0$ and all values of $\la$,
and reaches 2 for $p\simeq 1$, the intermediate $p$ values corresponding
to a crossover region.

The quantities $\rho^{-1}$ and $\Delta t$ are both related to the natural
timescale of the effective injection process. The main difference
between them is that $\rho^{-1}$ is effectively the rate of the
equivalent process, whereas $\De t$ is somehow a ``waiting time''
during which two injection events have little chance to occur
consecutively. $\Delta t$ is plotted in figure \ref{deltat}.

Figure \ref{rhoandrhodeltat} (a) shows as expected that the injection
process is very inefficient for $\la\simeq 0$ and also for $p\simeq
1$; obviously, this curve is qualitatively related to $\lan\eps\ran$,
since the efficiency of the injection has immediate consequences on
the mean injected power, but it is interesting to note that $\rho$ is
by no means constructed from $\lan\eps\ran$, but from cumulants of
higher order. 

Figure \ref{rhoandrhodeltat} (b) shows that the system
is really Poissonian in the regions $(\la\gtrsim0,p<1/2)$ and
$p\lesssim 1$, despite the fact that $\De t$ can be large
(fig. \ref{deltat}). Note also the vicinity of $\la=0$ and  $p>1/2$, 
where something interesting happens: two factors that elsewhere
favours the Poisson character of the process, namely $\la$ small and
$p>1/2$, are simultaneously at work here and act again each other. The
results of this ``collision'' is that the process is clearly not Poisson for
$p$ around $0.8$, due to huge values of $\De t$ (see
fig. \ref{deltat}); this is also a transitional region from a Poisson
process with one domain wall to a Poisson process with two domain
walls. 

Finally, when $p<1/2$ and $\la$ is away from zero (this is the region where the Fano fctor is larger than $1$),
 we find a
 non Poissonian process with $\De t$ of the order of $10\%$ to $15\%$ of
 $\rho^{-1}$. For instance, when $p=0$, $\rho$ increases as $\la$ increases:
 there is a crossover from a $\la$-limited regime to a bulk-limited regime for which a
 waiting time is observed. In this case, the probability of two close injection
 events is weakened because an injection event uses two spin flips:
 one flips of $s_0$ and then a flip of $s_1$ (for $p=0$, $s_1$ can
 flip only if $s_1=-s_0$); thus the probability of two events within
 $\de t$ goes like $\la^2(\de t)^4$ instead of $(\rho\de t)^2$ for a pure
 Poisson process (PPM). This explains the emergence of the waiting  time.

\section{Conclusion}

In this paper, we have presented a one-dimensional model of a
dissipative system, a half infinite chain of spins at $T=0$ in a
Glauber dynamics with a drift toward or away the boundary, sustained in a nontrivial stationary state by an
injection mechanism, namely the Poissonian flipping of the boundary
spin. We computed exactly the large deviation function of the injected
power and subsequently the first three cumulants of its probability distribution,
which account for the mean value of the injected power, its
fluctuations and the skewness of the fluctuations. Using a simple
phenomenological model and its refined version, we have shown that it
can account for the main physical
characteristics of the injection process very convincingly, allowing
for a relevant physical interpretation of the variations of the three cumulants
with the parameters $(\la,p)$ (rate of $s_0$ flipping, magnitude of
the drift), in terms of an effective rate of emission of energy
``quanta'', an average number of domain walls in each quantum,
and a possible waiting time after an injection event.

We can hope that this phenomenology could give an interesting scheme
to interpret some experiments, where the same kind of injection
mechanism is more or less reproduced. For instance, in a turbulent
experiment, unpinning of vortices created near a moving boundary could
 be a process of energy injection suitable for the description framework that we propose here.
  We can also think of the
bubble regime in the ebullition process, where the main part of the
energy transfer occurs via unpinning of vapour bubbles. We hope that some experimental results \cite{aumaitrefauve}
could find a simple
interpretation in the kinetic description that we give here.

Finally our mathematical calculations show that such simple out-of-equilibrium models are integrable: this opens the way
to more generalizations.

\section{Appendix: Fermionic approach to the time-integrated injected power}\label{calcul}

 It is useful to describe spin systems in the
dual representation of  domain walls : between the site $j$ and $j+1$ is
located the possible domain wall labelled $j$ for $j=-1,\ldots N$. The state
of the system is thus characterized by
$\mcal{C}=(n_{-1},\ldots,n_{N})$, where the
$n_i$ are either 0 (no domain wall) or 1. There are $2^{N+2}$ possible
states in this representation; note that the domain wall $n_{-1}$ does
not play any role, but is required to make the fermionic description tractable. The dynamical equation for the probability is given by
\begin{align}
  \pa_tP(\mcal{C})=\la [P(\mcal{C}_0)-P(\mcal{C})]+\sum_{j=1}^N
  [P(\mcal{C}_j)w(\mcal{C}_j\rightarrow \mcal{C})-P(\mcal{C})w(\mcal{C}\rightarrow \mcal{C}_j)]\label{preceq}
\end{align}
where $\mcal{C}_j$ holds for the state $\mcal{C}$ whose domain
wall variables $n_j$ and $n_{j-1}$ have been changed (according to
$n\rightarrow 1-n$). The $T=0$
asymmetric Glauber dynamics corresponds to
\begin{align}
  w(\mcal{C}_j\rightarrow \mcal{C})&=2[1-pn_j-qn_{j-1}]\\
  w(\mcal{C}\rightarrow \mcal{C}_j)&=2[pn_j+qn_{j-1}]
\end{align}
where $n_j$ and $n_{j-1}$ are the variables associated with the state
$\mcal{C}$ (we use this convention
hereafter), and $q=1-p$.

\medskip

We consider that each domain wall contributes as  an excitation 
 of energy $1$ to the global energy of the system.
We are interested in the energy $\Pi$ injected into the system up to time $t$ by the
Poissonian injection. Following \cite{derridalebowitz}, the route to this time integrated observable
begins with the consideration of the joint probability
$P(\mcal{C},\Pi,t)$, the probability for the system to be in the state
$\mcal{C}$ at time $t$ having received the energy $\Pi$ from the
injection. The dynamical equation for this quantity is readily
\begin{multline}
  \pa_tP(\mcal{C},\Pi)= \la
  \{P(\mcal{C}_0,\Pi-1)n_0+P(\mcal{C}_0,\Pi+1)(1-n_0) - P(\mcal{C},\Pi)\}\\
+\sum_{j=1}^N
  [P(\mcal{C}_j,\Pi)w(\mcal{C}_j\rightarrow \mcal{C})-P(\mcal{C},\Pi)w(\mcal{C}\rightarrow \mcal{C}_j)]
\end{multline}

We define next the  generating function of $\Pi$
as
\begin{align}
  F(\mcal{C})=\sum_{\Pi=-\infty}^{\infty}e^{\al\Pi}P(\mcal{C},\Pi)
\end{align}
This quantity, summed up over the states, yields the generating
function $\lan \exp(\al \Pi)\ran$ from which one derives its ldf
$g(\al)$:
\begin{align}
  \lan e^{\al \Pi}\ran\sur{\simeq}{t\rightarrow\infty}e^{t g(\al)}
\end{align}
This ldf $g(\al)$ is closely related to $f(p)$, the ldf of the probability
density function of $\Pi$, as they are Legendre transform of each
other \cite{farago1,farago2} :
\begin{align}
  \text{Prob}(\Pi/t=p)&\sur{\propto}{t\rightarrow\infty}\exp(tf(p))\\
f(p)&=\min_{\al}\left(g(\al)-\al p\right)
\end{align}

Let us write the dynamical equation for $F$ :
\begin{multline}\label{eqforF}
  \pa_t F(\mcal{C})=
\la\left[
  e^{\al}F(\mcal{C}_0)n_0+e^{-\al}F(\mcal{C}_0)(1-n_0)-F(\mcal{C})\right]\\
+2\sum_{j=1}^N [F(\mcal{C}_j)(1-pn_j-qn_{j-1})-F(\mcal{C})(pn_j+qn_{j-1})]
\end{multline}
The function $g(\al)$ can be expressed in terms of the
linear operator acting on the ``vector''  $[F(\mC)]_\mC$  in the r.h.s
of \myref{eqforF} :  it is in general its largest
eigenvalue.

Our problem belongs to the category of the
``free-fermions'' problems, for which a diagonalization of the
dynamics into independent ``modes'' can be achieved. The procedure is
described in \cite{faragopitard}, with references therein. In our
case, the operator in the r.h.s of equation \myref{eqforF} can be
turned into the following fermionic operator :
\begin{multline}
  H=\la[e^\al c_{-1}^\dag c_0^\dag+e^{-\al}c_0c_{-1}+e^\al c_0^\dag c_{-1}+e^{-\al}c_{-1}^\dag
  c_0-1]\\
+2\sum_{j=1}^N [c_jc_{j-1}+qc_j^\dag c_{j-1}+pc_{j-1}^\dag
  c_j-pc_j^\dag c_j-qc_{j-1}^\dag c_{j-1}]
\end{multline}
A symmetrisation procedure is a
prerequisite to solve the problem. We define a priori the change of
variables (note that the $\tilde{c}$ remain fermionic variables)
\begin{eqnarray}
c_{-1}=e^{-\al}\tilde{c}_{-1}&,&c^\dag_{-1}=e^{\al}\tilde{c}^\dag_{-1}\\
\forall j\geq1,  c_j=u_j\tilde{c_j}&,&
c_j^\dag=\frac{1}{u_j}\tilde{c}_j^\dag
\end{eqnarray}

where the $u_j$ are real quantities to be defined. The choice
\begin{align}
u_j&\sur{=}{j\geq 0}\left(\sqrt{\frac{q}{p}}\right)^j\equiv \nu^{j/2}
\end{align}

leads to the symmetrized expression (we omit the tildes immediately)
\begin{align}
  H&=\la[e^{2\al} c_{-1}^\dag c_0^\dag+e^{-2\al}c_0c_{-1}+ c_0^\dag c_{-1}+c_{-1}^\dag
  c_0-1]\nonumber\\
&\phantom{aaaaa}+2\sum_{j=1}^N [\nu^{j-1/2}c_jc_{j-1}+\sqrt{pq}c_j^\dag c_{j-1}+\sqrt{pq}c_{j-1}^\dag
  c_j-pc_j^\dag c_j-qc_{j-1}^\dag c_{j-1}]\\
&=\sum_{n,m=-1}^N\left[c^\dag_nA_{nm}c_m+\demi c_n^\dag B_{nm}c_m^\dag+\demi c_n
   D_{n,m}c_m\right]-\la\end{align}
where $A$ is a $(N+2)\times(N+2)$ tridiagonal, real and symmetric
  matrix, and $B$ and $D$ $(N+2)\times(N+2)$
   antisymmetric real; they are defined by   \begin{align}
A&=\left(
\begin{array}{ccccccc}
0 & \la & &&&& \\
\la & -2q  &2\sqrt{pq}&&&& \\
 & 2\sqrt{pq} & -2 &2\sqrt{pq} & &&\\
 &  & 2\sqrt{pq}& -2 & 2\sqrt{pq}  &&\\
& & &\ddots &\ddots &\ddots&  \\
&&&&2\sqrt{pq} &-2p
\end{array}\right)\\
B&=\left(
\begin{array}{ccccccc}
0 & \la e^{2\al} & &&&& \\
-\la e^{2\al} & 0  &&&&& \\
 &  & & & &&\\
 &  & &\mbox{\Huge 0}&  &&\\
& & & & &&  \\
\end{array}\right)\\
 D&=\left(
\begin{array}{ccccccc}
0 & -\la e^{-2\al} & &&&& \\
\la e^{-2\al} & 0  & -2\nu^\frac{1}{2}&&&& \\
 & 2\nu^\frac{1}{2} & 0 &-2\nu^\frac{3}{2} & &&\\
 &  & 2\nu^\frac{3}{2}& 0 & -2\nu^\frac{5}{2}  &&\\
& & &\ddots &\ddots &\ddots&  
\end{array}\right)  \end{align}
This Hamiltonian is diagonalizable, that is, it can be written
\begin{align}
  H&=
  \sum_q\La_q\left(\xi^\dag_q\xi_q-\demi\right)+\underbrace{\demi\text{Tr}A}_{-N}-\la
\end{align}
where the $\xi_q$ are fermionic operators linearly related to the
$c_j$ and the eigenvalues $\La_q$ are the eigenvalues with a positive
real part (we could have chosen the other half as well, see below) of the matrix
\begin{align}
  M_0&=\left(\begin{array}{lr} A& B\\D&-A\end{array}\right)
\end{align}
(The details of this procedure are exposed in \cite{faragopitard};
note that the lack of translational invariance prevents the use of
a Fourier transformation).

The eigenvalues of $H$ are thus given by
\begin{align}
  \demi\sum_q\La_q\eps_q-N-\la
\end{align}
where the $\eps_q$ are $\pm 1$. In particular, the largest eigenvalue of $H$ reads
\begin{align}\label{formalg}
  g(\al)&=\demi\sum_q\text{Re}(\La_q)-N-\la\\
  &=\frac{1}{4i\pi}\oint d\mu \mu \frac{\chi'_0(\mu)}{\chi_0(\mu)}-N-\la\label{gintegral}
\end{align}
where $\chi_0$ is the characteristic polynomial of $M_0$ and the
contour of integration is diverging half circle leant on the imaginary
axis, with its curved part pointing toward the region $\text{Re}(\mu)>0$.

\subsection{The characteristic polynomial : introduction}

The problem is now equivalent to finding the characteristic polynomial
of $M_0$. We can take advantage of the emptiness of $B$.
We define $E(\mu)=(A+\mu)^{-1}D(A-\mu)^{-1}$. Multiplying
 $\mu\text{Id}-M_0$ by
 \begin{align}
   \left(
   \begin{array}{lr}
     (\mu-A)^{-1}&0\\ (\mu+A)^{-1}D(\mu-A)^{-1}&(\mu+A)^{-1}
   \end{array}\right)
 \end{align}
we see that
\begin{align}\label{chi0}
  \chi_0(\mu)&=\chi_A(\mu)\chi_A(-\mu)\det(1+BE(\mu))
\end{align}
where $\chi_A(\mu)=\det(A-\mu)$.
Besides,
\begin{align}
\det(1+BE)&=(1-\la e^{2\al} E_{-1,0})(1+\la e^{2\al} E_{0,-1})+\la^2e^{4\al} E_{0,0}E_{-1,-1}\\
&=[1+\la e^{2\al} E_{0,-1}(-\mu)][1+\la e^{2\al}E_{0,-1}(\mu)]+\la^2 e^{4\al} E_{0,0}(\mu)E_{-1,-1}(\mu)
\end{align}
where we exploited the fact that $E^T(\mu)=-E(-\mu)$. Note in passing
that the symmetry of this expression with respect to
$\mu\rightarrow-\mu$, is here explicit,  as the $E_{jj}$ are antisymmetric functions
of $\mu$.

\subsection{Some minors of $\mu\text{Id}+A$}

We term $\De_j$, ($j=0,\ldots,N+1$) the determinant of the minor of $(\mu+A)$ obtained by
keeping the $(N+1-j)\times(N+1-j)$ matrix located at the bottom right
side of $(\mu+A)$ (one adopts the convention $\De_{N+1}=1$). Note that
$\De_N=-2p+\mu$, $\De_{N-1}=(\mu-2)(\mu-2p)-4pq$, and that we have
that $\det(\mu+A)=\mu\De_0-\la^2\De_1$. We have also an explicit
formula, valid for $j\neq 0$:
\begin{align}
  \De_j&=\frac{1}{1-4pq x_+^2}\left[(2qx_++1)x_+^{j-N-1}-2qx_+(2px_++1)(4pqx_+)^{-j+N+1}\intvidepetit\right]\label{De}
\end{align}
where $x_+$ is conventionally the root of the polynomial
$-4pqX^2+(\mu-2)X-1=0$ \textit{with the largest modulus}. Note that
$x_+$ and $\De_j$ depends on $\mu$. Later on, we will denote
$x_-=x_+(-\mu)$ and $\De_j^-=\De_j(-\mu)$; let us stress here that $x_+$
and $x_-$ are roots of different polynomials.

\subsection{The characteristic polynomial : explicit calculation}

From the definition $E=(A+\mu)^{-1}D(A-\,u)^{-1}$, one gets after some computations
\begin{align}
W&=\frac{1}{2q}\sum_{j=1}^N(4q^2)^j[\De_{j+1}\De_j^--\De_{j+1}^-\De_j]\label{W}\\
  E_{0,-1}&=\frac{\la}{\det(A+\mu)\det(A-\mu)}\left[\intvidepetit
  e^{-2\al}\De_1(\mu\De_0^--\la^2\De_1^-)+2\mu W\right]\\
E_{-1,-1}&=\frac{-\la^2}{\det(A+\mu)\det(A-\mu)}\left[\intvidepetit
  (\De_1\De_0^--\De_1^-\De_0)e^{-2\al}+2W\right]\\
E_{0,0}&=\frac{-\mu}{\det(A+\mu)\det(A-\mu)}\left[\intvidepetit
  2\la^2e^{-2\al}\De_1\De_1^--2\mu W\right]
\end{align}
whence one deduces
\begin{align}
 \chi_0(\mu)&=
 -\mu^2\De_0\De_0^-+2\la^2\mu[\De_0^-\De_1-\De_0\De_1^-]+4\la^2\mu e^{2\al}W
\end{align}

Let us analyse $W$. Using equations \myref{W} and \myref{De}, one
easily shows that $W$ has 5 different terms, respectively proportional
to $((4pq)^2x_+x_-)^N$, $(x_+x_-)^{-N}$, $(x_+/4pqx_-)^N$,
$(x_-/4pqx_+)^N$, and $(4q^2)^N$. From the definition of $x_+$ and
$x_-$, one has always $4pq|x_\pm|^2>1$. This shows that the first term always
dominates all but the last. A slight issue arises here, for the last
is not always the least: for $q<1/2$, the first is still dominating,
but for $q>1/2$ this is not the case for all values of $\mu$ in the
right half plane $\mcal{P}=\{\mu;$~Re($\mu)\geq0\}$. Let us term $\mcal{J}$ the
zone in $\mcal{P}$ where $4q^2\geq(4pq)^2|x_+x_-|$. Two key features of
$\mcal{J}$ are that (i) it is bounded (compact) (ii) it crosses the
vertical line Re(~$\mu)=0$ only at one point, $\mu=0$. To prove that,
we remark that on that line, $x_-=x_+^*$ (complex conjugate); moreover
$x_+(\mu=0)=1/2p$ and $x_+(-iy)=x_+(iy)^*$: the maximum principle
leads to the conclusion that $|x_+(iy)|$ is a function of $y$, minimum
at $y=0$.

As a result, the contour of integration in equation \myref{gintegral}
can always be chosen such that, except for the single point $\mu=0$,  it does not cross the region $\mcal{J}$
(it  encloses it anyway). In that case, the thermodynamic limit can be
safely taken for all values of $q$, and leads to the \textit{complete
  vanishing} of the term proportional to $(4q^2)^N$ in the result,
dominated by the first one. This mathematical argument yields a great
simplification, as one can consider that at the thermodynamic limit, $W$
is always dominated by the term $\propto [(4pq)^2x_+x_-]^N$ and throw
away the others.

\medskip

According to the preceding discussion, we are left with
\begin{align}
  W&\sur{=}{N\rightarrow\infty}\frac{2pA_+A_-}{4p^2x_+x_--1}(16p^2q^2x_+x_-)^N(x_--x_+)\\
A_\pm&=\frac{2qx_\pm(2px_\pm+1)}{4pqx_\pm^2-1}\\
\end{align}
Similarly, we can write for $j\geq 1$
\begin{align}
  \De_j&\sur{=}{N\rightarrow\infty}A_+(4pqx_+)^{-j+N+1}
\end{align}
As regards $\De_0$, we have $\De_0=(-2q+\mu)\De_1-4pq\De_2$. Thus, 
\begin{align}
  \De_0&\sur{=}{N\rightarrow\infty}A_+(4pqx_+)^N[-2q+\mu-1/x_+]\\
&=A_+(4pqx_+)^N\times 2p(1+2qx_+)
\end{align}
As a result, we get
\begin{align}
  \chi_0(\mu)&=A_+A_-((4pq)^2x_+x_-)^N\nonumber\\
&\phantom{aaaaaaaa}\times\left(\intvide-4p^2\mu^2(1+2qx_+)(1+2qx_-)+8pq\la^2\mu(x_--x_+)+8p\la^2\mu
  e^{2\al}\frac{x_--x_+}{4p^2x_+x_--1}\right)\\
&=A_+A_-((4pq)^2x_+x_-)^N 4p\mu\nonumber\\
&\phantom{aa}\times\left(\intvide-p\mu(1+2qx_+)(1+2qx_-)+2\la^2(x_--x_+)\frac{4p^2qx_+x_-+p}{4p^2x_+x_--1}+2\la^2(x_--x_+)\frac{e^{2\al}-1}{4p^2x_+x_--1}\right)
\end{align}
This expression can be transformed in the following way: we can
demonstrate the relations
\begin{align}
  \frac{1}{4pqx_+x_--1}&=\frac{1}{2\mu}\left(\frac{1}{x_-}-\frac{1}{x_+}\right)\\
(1+2qx_+)(1+2qx_-)(4p^2x_+x_--1)&=\mu^2x_+x_-\frac{1+4pqx_-x_+}{1-4pqx_-x_+}
\end{align}
(for the first, multiply the lhs by $(x_+/x_--1)^{-1}$; for the second, use
the fact that $x_\pm$ are roots of $2^\text{nd}$ degree polynomials). 
Thus we can write
\begin{align}
\chi_0(\mu)&=A_+A_-((4pq)^2x_+x_-)^N 4p^2\nonumber\\
&\phantom{aa}\times\left(\intvide
\left(4\la^2-\mu^2\right)(1+2qx_+)(1+2qx_-)+\frac{2\la^2\mu}{p}(x_--x_+)\frac{e^{2\al}-1}{4p^2x_+x_--1}\right)\\
&=A_+A_-((4pq)^2x_+x_-)^N 4p^2\left(4\la^2-\mu^2\right)(1+2qx_+)(1+2qx_-)\nonumber\\
&\phantom{aa}\times\left(\intvide
1+\frac{4\la^2/p}{4\la^2-\mu^2}\frac{e^{2\al}-1}{4pqx_+x_-+1}\right)\label{finalchi0}
\end{align}

\subsection{An integral formula for $g$}

It is  useful for the sequel to give explicit formulas for $x_+$
and $x_-$ in the half plane   $\mcal{P}=\{\mu;\text{Re}(\mu)\geq 0\}$.
A careful inspection shows that $x_+$ is given by
\begin{align}
  x_+(\mu)&=\left\{
  \begin{array}{ll}
    (8pq)^{-1}\times\left(\intvidepetit
    \mu-2-\sqrt{(\mu-2)^2-16pq}\right)&\text{if Re$(\mu)\in [0,2]$}\\
    (8pq)^{-1}\times\left(\intvidepetit
    \mu-2+\sqrt{(\mu-2)^2-16pq}\right)&\text{if Re$(\mu)>2$}
  \end{array}\right.
\end{align}
It must noted that, contrary to appearances, $x_+$ is analytic on the
line $\text{Re}(\mu)=2$. It has however anyway a branch cut,  localized on the segment
$\mu\in[2-4\sqrt{pq},2+4\sqrt{pq}]$.

The behaviour of $x_-$ is entirely different:
\begin{align}
  x_-(\mu)&=-(8pq)^{-1}\times\left(\intvidepetit
    \mu+2+\sqrt{(\mu+2)^2-16pq}\right)
\end{align}
and is analytic on $\mcal{P}$.

\bigskip

Let us go back to formula \myref{gintegral}. We see that only  the
logarithmic derivative of $\chi_0$  is involved, so we can
handle the different terms of \myref{finalchi0} separately. For sake
of clarity, we define
\begin{align}
  I[f]=\frac{1}{4i\pi}\oint_+ d\mu \mu \frac{f'(\mu)}{f(\mu)}
\end{align}
over a contour (followed counterclockwise) in $\mcal{P}$ large enough to encircle all the
singularities of $f$.
\begin{itemize}
\item the terms $x_\pm^N$: as  $x_-$ it is an analytic
  function of $\mu$ over $\mcal{P}$, we get $I[x_-^N]=0$ (obviously
  neither $x_-$ nor $x_+$ can go to zero). The term $I[x_+^N]$ gives a
  nonzero contribution due to the branch cut of $x_+$. We remark that
  on it, $|x_+|=2\sqrt{pq}$ and $x_+$ describes counterclockwise the
  circle of radius $2\sqrt{pq}$. Thus,
  \begin{align}\label{transfo}
    I[x_+^N]&=\frac{N}{4i\pi}\oint_{|z|=2\sqrt{pq}}\frac{dz}{z}(4pqz+z^{-1}+2)=N
  \end{align}
\item The term $(4\la^2-\mu^2)$ yields $I[4\la^2-\mu^2]=\la$.
\item We show easily that $A_\pm(1+2qx_\pm)=\mp \mu x_\pm^2/(4pq
  x_\pm^2-1)$. We conclude easily that $I[A_-(1+2qx_-)]=0$, for
  $4pqx_-^2-1$ never vanishes. As regards $I[A_+(1+2qx_+)]$,  a
  transformation similar to \myref{transfo} gives also
  $I[A_+(1+2qx_+)]=0$.
\end{itemize}

Finally, from these results, we see that all terms but the last in
\myref{finalchi0} cancel with constant terms in  \myref{gintegral}. To
give the final result a convenient form we remark that the contour of
integration can be make infinite, that the semicircular part gives a
vanishing contribution to the result, and that on the vertical line
$\text{Re}(\mu)=0$, $x_+=x_-^*$. We can thus write
\begin{align}
  \boxed{g(\al)=\frac{2}{\pi}\int_{0}^{\infty} dy 
  \log\left(1+\frac{\la^2/4p}{\la^2/4+y^2}\frac{e^{2\al}-1}{\psi(y)+1}\right)}\label{finalformulaforg}\\
\psi(y)=4pq|x_-(4iy)|^2=(4pq)^{-1}\times \left|2iy+1+\sqrt{(2iy+1)^2-4pq}\right|^2
\end{align}

We verify immediately that $g(0)=0$ as expected. We can also check
that $g'(0)=\lan\Pi\ran$:
\begin{align}
  g'(0)&=\frac{2\la^2}{p\pi}\int_{-\infty}^\infty
  \frac{dy}{(\la^2+4y^2)(\psi(y)+1)}\\
&=\frac{-\la^2}{2p\pi}\int_{-\infty}^\infty
  \frac{dy}{\la^2+4y^2}\left(\frac{1}{x_+(4iy)}+\frac{1}{x_-(4iy)}\right)\\
&=-\frac{\la^2}{p\pi}\text{Re}\int_{-\infty}^\infty
  \frac{dy}{\la^2+4y^2}\frac{1}{x_-(4iy)}=\frac{\la}{2p}(\la+1-\sqrt{(\la+1)^2-4pq})
\end{align}

We mention (without computations) also the result we would obtain, if we had 
considered, like in \cite{faragopitard}, two half lines of spins connected to
$s_0$ (i.e. spins numbered $s_{-1},s_{-2},\ldots$). In this case,
despite the fact that the two subsystems are connected only via the
Poisson spin $s_0$, they are nontrivially coupled to each other, and
the corresponding large deviation function of the cumulants reads
\begin{align}
  g(\al)&=\frac{2}{\pi}\int_0^\infty dy \log\left(1+\frac{\la^2/4p}{\la^2/4+y^2}\frac{e^{2\al}-1}{\psi(y)+1}\left[2+\frac{e^{2\al}-1}{p(\psi(y)+1)}\right]\right)
\end{align}
We see clearly that there is no simple correspondence between the half
line model and the two half lines model: $g$ is multiplied inside the
logarithm by an $\al$-dependent term.

 \end{document}